# Two-dimensional vector solitons stabilized by a linear or nonlinear lattice acting in one component

Olga V. Borovkova,[1] Boris A. Malomed,[2] and Yaroslav V. Kartashov [1]

[1]*ICFO-Institut de Ciencies Fotoniques, and Universitat Politecnica de Catalunya, Mediterranean Technology Park, 08860 Castelldefels (Barcelona), Spain*

[2]*Department of Physical Electronics, School of Electrical Engineering, Faculty of Engineering, Tel Aviv University, Tel Aviv 69978, Israel*

The subject of the work is the stabilization of two-dimensional (2D) two-component (vector) solitons, in media with the attractive cubic nonlinearity, against the collapse by a linear lattice (LL, which is induced by a periodic modulation of the refractive index in optics, or created as an optical lattice in BEC), or by a nonlinear lattice (NL, induced by a periodic modulation of the nonlinearity coefficient). We demonstrate that, due to the XPM (cross-phase-modulation) coupling between the components, the LL or NL acting on a *single* component is sufficient for the stabilization of vector solitons, that include a component for which the self-focusing medium is uniform. In the case of the LL, the vector solitons are stable almost in their entire existence domain, while the NL can only stabilize the solitons in which the component affected by the lattice carries the norm which is comparable to, or larger than the norm of the component in the uniform medium.

Spatial solitons, such as self-trapped optical beams, may exist in a variety of nonlinear media. The crucial issue is the stability of such solitons, especially in case of two transverse dimensions. In particular, the formation of solitons in the medium with the Kerr (cubic self-focusing) nonlinearity is a well-known challenge, because collapse is an inherent feature of 2D settings with the same nonlinearity [1]. Various strategies were developed for the stabilization of such 2D modes. These include the creation of linear lattices (LLs), in the form of a periodic transverse modulation of the refractive index of optical materials, or optical lattices in BEC (Bose-Einstein condensates) [2-7]. The LLs can stabilize not only the simplest scalar (single-component) fundamental solitons, but also vortex and multipole solitons (see reviews [1,8,9] and references therein). Another stabilization mechanism relies on the transverse modulation of the nonlinearity coefficient, i.e., the nonlinear lattice (NL) [10]. In 1D settings, scalar solitons supported by NLs [10-13], as well as by combinations of linear and nonlinear lattices [10,14-18], have been studied in detail. A much more challenging issue is the stabilization of 2D solitons by the modulation of the nonlinearity [19-22]. Only recently it was shown that this is possible in structures with sharp variations of the nonlinearity coefficient – in particular, in arrays of Kerr-nonlinear circles ("cylinders") embedded into a linear [21] or saturable [22] host medium.

The solitons may include two components (optical beams with different polarizations or carrier wavelengths, or a mixture of different atomic states in BEC), that are coupled by the XPM nonlinearity into vector states. The XPM coupling greatly enriches the variety of soliton families and affects their stability [23-29]. While it was demonstrated that LLs readily support stable 2D vector solitons [30,31], only 1D vector modes were considered, up to date, in NLs [32,33]. In all the cases studied before, it was also assumed that lattice acted equally on *both* components of vector systems. 2D vector solitons in NLs were not considered as yet at all. In particular, a challenging issue is whether such solitons can be made stable when the lattice (linear or nonlinear) selectively acts on a *single* component, the medium being uniform for the other one.

In this Letter, we address the existence and stability of 2D two-component bright solitons, in the case of the linear or nonlinear lattice affecting only one component in the system. We demonstrate that the XPM coupling may result in the stabilization of the entire vector complex, despite the fact that the second component is not directly stabilized by any lattice (i.e., such a scalar mode would decay or collapse). We conclude that the "single-sided" LL, which acts on a single component, can stabilize vector modes almost in their entire existence domain (even when the component affected by the lattice carries a much smaller norm than the second component), while the stabilization by the single-sided NL requires that the norm of the component affected by the lattice should be comparable to or larger than the norm of the other component.

The 2D evolution of coupled light beams, or wave functions describing the mixture of different atomic species in the BEC, in the presence of the single-sided linear or nonlinear lattice, is described by coupled nonlinear Schrödinger/Gross-Pitaevskii equations for scaled field amplitudes/wave functions $q_{1,2}$:

$$
\begin{aligned}
i\frac{\partial q_1}{\partial \xi} &= -\frac{1}{2}\left(\frac{\partial^2 q_1}{\partial \eta^2} + \frac{\partial^2 q_1}{\partial \zeta^2}\right) - q_1[\sigma(\eta,\zeta)|q_1|^2 + C|q_2|^2] - pR(\eta,\zeta)q_1, \\
i\frac{\partial q_2}{\partial \xi} &= -\frac{1}{2}\left(\frac{\partial^2 q_2}{\partial \eta^2} + \frac{\partial^2 q_2}{\partial \zeta^2}\right) - q_2(|q_2|^2 + C|q_1|^2),
\end{aligned}
\tag{1}
$$

where $\xi$ is the normalized propagation distance (time in the case of the BEC) and $\eta, \zeta$ are the transverse coordinates. Function $\sigma(\eta,\zeta)$ describes the nonlinearity landscape in the case when component $q_1$ is subject to the action of the NL, while function $R(\eta,\zeta)$ represents the LL with depth $p$, and $C$ is the XPM coefficient. To analyze effects of the NL, we set $p = 0$ and adopt the nonlinearity-modulation landscape with $\sigma = 1$ inside a square array of circles if radius $w_{\mathrm{r}}$, with spacing $w_{\mathrm{s}}$ between them, while between the circles the medium is linear, with $\sigma = 0$. The results are reported here for $w_{\mathrm{r}} = 0.5$ and $w_{\mathrm{s}} = 2$, which adequately represents the generic situation. We stress that the nonlinearity acting on the second component, $q_2$, is spatially uniform. In contrast, to consider LL effects, we assume that the nonlinearity is uniform in both components, i.e., $\sigma \equiv 1$, but the linear potential is present, with $p \neq 0$ and $R(\eta,\zeta) = \cos(\Omega\eta)\cos(\Omega\zeta)$. In this paper, we set $p = 4$ and $\Omega = 2$.

Crucially important for the stability of vector solitons in the present models is the value of the XPM coefficient, $C$, which depends on physical settings. In optical systems, $C = 1$ for mutually incoherent light beams, $C = 2/3$ for coherent beams with orthogonal linear polarizations in strongly birefringent media, and $C = 2$ for circular polarizations, or for two waves with different carrier wavelengths. In addition to that, $C$ may be large in organic materials, and it may be engineered in photonic-crystal fibers. Additional possibilities for the manipulation of the XPM coupling are offered by photorefractive crystals, through varying the polarization of the beams and/or elements of the underlying electro-optic tensor. Therefore, we consider $C$ as a control parameters varying in a relatively broad range.

As concerns the distinctive feature of the present models, *viz.*, that the lattice acts on the single component, it can be easily implemented in optics, using the lattice built of alternating layers of isotropic and birefringent materials, with one of the two values of the refractive index in the birefringent material matched to that of the isotropic medium. The polarization component pertaining to this common values of the refractive index will not feel the action of the lattice, while the orthogonal polarization component will be subject to the refractive-index modulation. In the BEC, the selectively acting optical lattice may be realized using a mixture of two bosonic isotopes, with nearly equal atomic masses, the optical fields which build the corresponding LL being near-resonant for one species only. On the other hand, the selective NL in BEC can be easily realized by means of the accordingly tuned Feshbach resonance, cf. Refs. [10-13].

We search for vector-soliton solutions to Eq. (1) as $q_{1,2}(\eta, \zeta, \xi) = w_{1,2}(\eta, \zeta) \exp(ib_{1,2}\xi)$, where $b_{1,2}$ are the corresponding propagation constants. We are looking for simplest solutions with the fundamental (bell-shaped) intensity distributions in both components. Nevertheless, because the lattice acts on one component, the shapes of $w_1$ and $w_2$ are always different, even if $b_1 = b_2$ (in contrast to the uniform cubic medium, where one has $w_1 = \text{const} \cdot w_2$ for $b_1 = b_2$). To cope with the large number of parameters in the model, we fix the propagation constant of the first component, $b_1 = 3$, and vary $b_2$ and $C$. A detailed numerical analysis indicates that the results do not change significantly for other values of $b_1$, as long as they belong to domain where corresponding scalar soliton, with $q_1 \neq 0$, $q_2 = 0$, supported by NL or LL is *stable*. Notice also that Eqs. (1) conserve the total and partial norms of the two components (the energy flows in optics, or numbers of atoms in BEC):

$$U = U_1 + U_2 \equiv \int \int_{-\infty}^{\infty} (|q_1|^2 + |q_2|^2) d\eta d\zeta, \qquad (2)$$

which also determine the power sharing between the components, $S_{1,2} \equiv U_{1,2} / U$.

Typical field profiles of 2D vector solitons are shown in Fig. 1, by means of cross sections drawn through $\zeta = 0$. As said above, in both cases of the nonlinear [Figs. 1(a) and 1(b)] and linear [Figs. 1(c) and 1(d)] lattices, the propagation constant $b_1 = 3$ was selected so that, in the absence of the second component $q_2$ for which the medium is uniform, the corresponding scalar soliton with $q_1 \neq 0$, $q_2 = 0$, $C = 0$ is stable (for a detailed analysis of the stability of scalar solitons in NLs see Ref. [21]). Only in this case the stabilization of the second component may be provided by XPM-induced coupling to its counterpart which is directly stabilized by the NL. At $C > 1$, the increase of $b_2$ results in a decrease of the amplitude of the second component, in both cases of the LL and NL, so that for sufficiently large $b_2$ the second component becomes much weaker than the first one [see Figs. 1(b) and 1(d)], and, at certain value $b_2 = b_2^{\text{upp}}$, the second component vanishes altogether, so that the vec-tor soliton degenerates into a scalar one with $w_1 \neq 0$, $w_2 = 0$. As said above, the resulting scalar soliton is stable because of the proper choice of $b_1$. In contrast, the decrease of $b_2$ results in a reduction of the first component [see Figs. 1(a) and 1(c)], so that, at another threshold, $b_2 = b_2^{\text{low}}$, the vector soliton degenerates into a different scalar mode, with $w_1 = 0$, $w_2 \neq 0$. Since for second component the medium is uniform, the latter scalar mode is tantamount to the usual Townes soliton in the uniform Kerr medium, which is always unstable [1]. Therefore, in the present model the vector solitons exist in a limited range of values of the propagation constant, $b_2^{\text{low}} \leq b_2 \leq b_2^{\text{upp}}$. Naturally, this existence domain dramatically depends on the XPM coefficient, $C$.

Remarkably, the situation described above for $C > 1$ is completely reversed at $C < 1$. In this case, the norm of the second component increases when $b_2 \to b_2^{\text{upp}}$ and vanishes when $b_2 \to b_2^{\text{low}}$. In all cases, the shapes of both the first and second components are smooth, without pronounced oscillations that might be induced by the lattice. This is because the second component is not affected by the lattice, while the propagation constant of the first component is sufficiently large to ensure its stability in the scalar case, which implies that the corresponding field $w_1$ is well localized within few sites of the LL or NL.

Properties of the vector solitons are summarized in Fig. 2 that shows typical dependencies of the norm and energy sharing between the soliton's components versus $b_2$, for both the NL- [Figs. 2(a) and 2(b)] and LL-based [Figs. 2(c) and 2(d)] models. The norm is a nonmonotonous function of $b_2$ - at $C = 2$, it first decreases with the increase of $b_2$, but then it starts to grow, see Figs. 2(a) and 2(c). The behavior of the total norm is different at $C < 1$ - in that case, $U$ first grows and then diminishes with the increase of $b_2$. Since at $C = 2$ the norm of the second component grows when $b_2 \to b_2^{\text{low}}$, while the norm of the first component increases at $b_2 \to b_2^{\text{upp}}$, the energy sharing $S_{1,2}$ between the soliton components drastically varies in interval $[b_2^{\text{low}}, b_2^{\text{upp}}]$, as shown in Figs. 2(b) and 2(d). Interestingly, in the case of the NL the values of the total norm of the vector solitons at the borders of the existence domain (i.e., at $b_2 \to b_2^{\text{low}}$ and $b_2 \to b_2^{\text{upp}}$) are very similar, despite the fact that the energy (norm) is concentrated in different components. However, in the presence of the LL, $U(b_2 = b_2^{\text{upp}})$ is considerably smaller than $U(b_2 = b_2^{\text{low}})$.

The transformation of the vector soliton into a stable scalar mode at $b_2 \to b_2^{\text{upp}}$, and its transformation into the unstable Townes soliton at $b_2 \to b_2^{\text{low}}$ (recall that this picture holds for $C > 1$, while for $C < 1$ the situation is inverse) suggests that the vector state may be stable in a range of values of $b_2$ close to $b_2^{\text{upp}}$, and the stability is lost somewhere inside the existence domain $[b_2^{\text{low}}, b_2^{\text{upp}}]$.

The main result reported in this Letter is that the stabilization of the vector solitons is indeed possible for both cases of the LL and NL acting on the *single* component. Figure 3 shows the domains of the existence and stability for the vector solitons in the plane of $(C, b_2)$. The stability domain was obtained by collecting simulation results for the propagation of perturbed vector solitons corresponding to different values of $b_2$ and $C$, over a huge distance, $\xi \sim 1000$. The existence domains of the vector solitons shrink around $C = 1$, but then expand with the increase or decrease of the XPM strength. The structure of the existence domains for both the NL [Fig. 3(a)] and LL [Fig. 3(b)] is similar. The vector solitons exist between black and green curves, with the second component vanishing on the green curve, and first component vanishing along the black one. The stability domains, marked by "s", are located close to the green curves, while instability domains, marked by "u", are situated close to the black curves. The border of the stability domain is depicted by the red curve. One can see from Fig. 3(b) that the LL stabilizes the vector solitons almost in their entire existence domain, while for the NL the widths of the stability and instability domains in Fig. 3(a) are comparable. The main reason behind this fact is that the NL strength, hence its ability to stabilize the solitons, depends on the power of the soliton itself. Therefore, in such lattices the amplitude of the first component, which is subject to the action of the lattice, should be comparable with or larger than the amplitude of the second component. This is in contrast to the case of the LL, where even weak $w_1$ carrying a relatively small norm, can result in the stabilization of the *entire* vector complex. Figure 3(c) shows the ratio of norms $U_2/U_1$ of the components of the vector soliton, supported by the NL and taken at a border of the stability domain [the red curve in Fig. 3(a)], versus XPM coefficient $C$ . One can clearly see that, in the case of the NL, this ratio is of the order of unity for any value of $C$ in the range of $[0.2, 2.0]$. The ratio $U_2/U_1$ apparently decreases as one moves deeper into the stability domain by varying $b_2$ . In contrast to that, in the case of the LL, when the stability domain almost entirely overlaps with the entire existence domain, the ratio $U_2/U_1$ may take on much higher values for stable vector solitons. In particular, for $C = 2$ the ratio at the stability border in the model with LL is $U_2/U_1 = 5.5$ , while for $C = 0.5$ this ratio is $U_2/U_1 = 17.6$ . These results clearly indicate that even a small-amplitude components affected by the LL can readily stabilize the entire vector soliton.

The results reported here do not vary dramatically upon the variation of the propagation constant of the first component, $b_1$ , although the variation of $b_1$ considerably affects the solitons' existence and stability domains. In particular, the increase of $b_1$ at $C > 1$ results in a considerable expansion of the existence and stability domains. For instance, at $b_1 = 10$ and $C = 2$ , the vector solitons in the nonlinear lattice exist between $b_2^{\text{low}} = 3$ and $b_2^{\text{upp}} = 34.5$ , and become completely stable at $b_2 > 10.5$ .

Typical examples of the stable propagation of the vector solitons perturbed by a strong white noise in the models with both the NL and LL are shown in Figs. 4(a) and 4(b), respectively. While unstable vector solitons usually decay, the stable ones propagate without any considerable distortion over indefinitely long distances.

Summarizing, we have shown that the XPM coupling between two components of the vector solitons, in systems with only one component subject to the stabilizing action of the linear or nonlinear lattice (LL or NL), readily results in the stabilization of the vectorial soliton complexes. While even a weak component affected by the LL is sufficient for the stabilization of the entire complex, the stabilization in the NL requires that the norm in the component which is directly supported by the lattice should be comparable to or greater than the norm of other component, for which the medium is uniform.

**Figure captions**

Figure 1.   (Color online) Field distributions $w_{1,2}(\eta)$ at $\zeta = 0$ in 2D vector solitons with (a),(b) nonlinear and (c),(d) linear lattices acting on the first component only. In all cases, $C = 2$, $b_1 = 3$ are fixed, while $b_2 = 1$ (a), 9 (b), 0.4 (c), and 2.5 (d). Solitons shown in panels (a),(c) are unstable, while solitons from panels (b),(d) are stable.

Figure 2.   (Color online) The total energy flow (a) and power sharing between the two components (b), versus $b_2$ at $b_1 = 3$ and $C = 2$, for the vector soliton, with the nonlinear lattice acting on the first component. Panels (c) and (d) show the same, but when the linear lattice acts in the first component. Points correspond to solitons shown in Fig. 1.

Figure 3.   (Color online) Domains of the stability ("s") and instability ("u") for vector solitons, with (a) nonlinear and (b) linear lattice acting on the first component, in the $(C, b_2)$ plane for $b_1 = 3$. In both cases, the vector solitons exist at $b_2^{\mathrm{low}} \leq b_2 \leq b_2^{\mathrm{upp}}$, in the region between the black and green curves. (c) The ratio of norms (energy flows ) $U_2/U_1$ of the two components of the vector soliton, taken at the border of the stability domain depicted in panel (a), versus $C$.

Figure 4.   (Color online) The stable propagation of a perturbed vector soliton with the nonlinear (a) and linear (b) lattice acting in the first component, at $C = 2$. Only the distribution of the absolute value of the field in the second component is shown. The solitons displayed in (a) and (b) corresponds to $b_1 = 3$, $b_2 = 5.5$, and $b_1 = 3$, $b_2 = 2.5$, respectively.

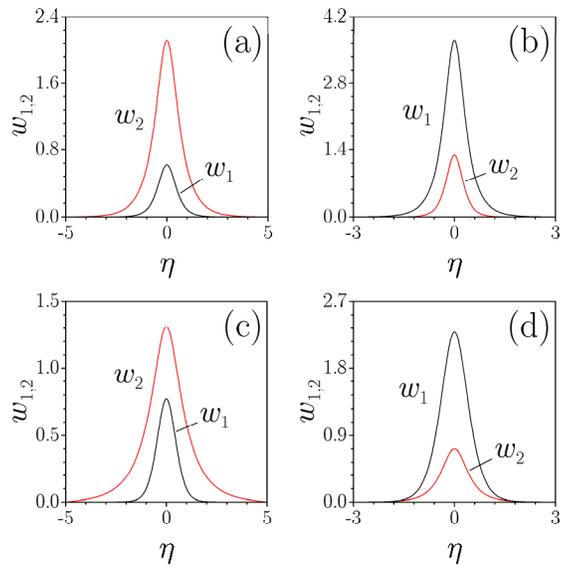

Figure 1. (Color online) Field distributions $w_{1,2}(\eta)$ at $\zeta = 0$ in 2D vector solitons with (a),(b) nonlinear and (c),(d) linear lattices acting on the first component only. In all cases, $C = 2$, $b_1 = 3$ are fixed, while $b_2 = 1$ (a), 9 (b), 0.4 (c), and 2.5 (d). Solitons shown in panels (a),(c) are unstable, while solitons from panels (b),(d) are stable.

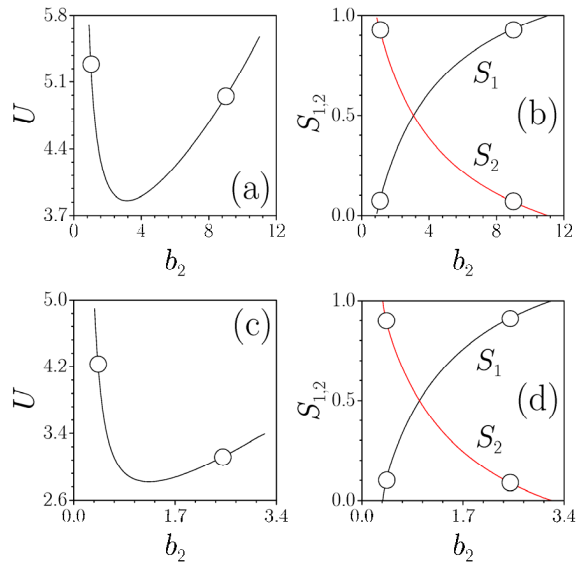

Figure 2.   (Color online) The total energy flow (a) and power sharing between the two components (b), versus $b_2$ at $b_1 = 3$ and $C = 2$, for the vector soliton, with the nonlinear lattice acting on the first component. Panels (c) and (d) show the same, but when the linear lattice acts in the first component. Points correspond to solitons shown in Fig. 1.

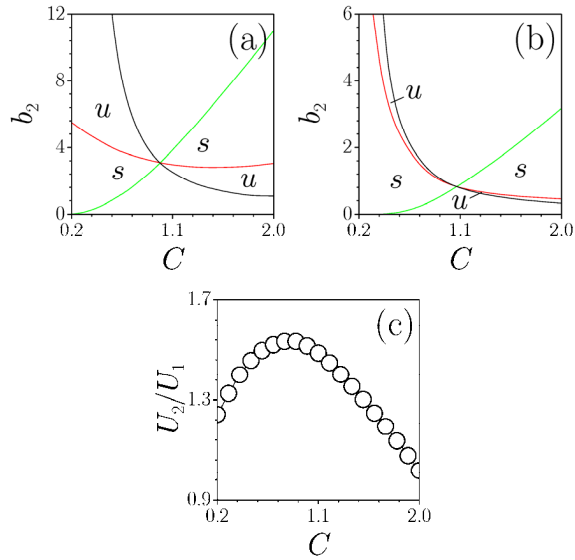

Figure 3.    (Color online) Domains of the stability ("s") and instability ("u") for vector solitons, with (a) nonlinear and (b) linear lattice acting on the first component, in the $(C, b_2)$ plane for $b_1 = 3$. In both cases, the vector solitons exist at $b_2^{\text{low}} \leq b_2 \leq b_2^{\text{upp}}$, in the region between the black and green curves. (c) The ratio of norms (energy flows ) $U_2/U_1$ of the two components of the vector soliton, taken at the border of the stability domain depicted in panel (a), versus $C$ .

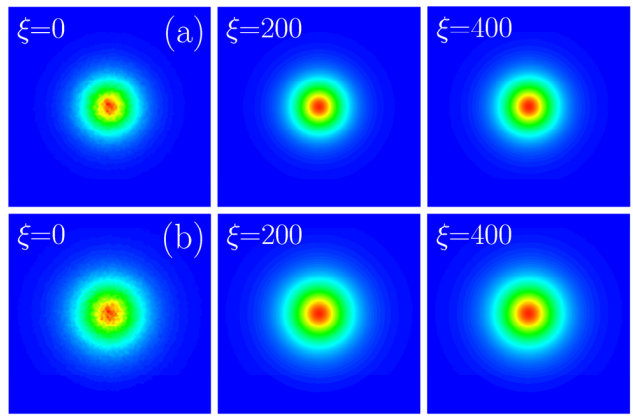

Figure 4. (Color online) The stable propagation of a perturbed vector soliton with the nonlinear (a) and linear (b) lattice acting in the first component, at $C = 2$. Only the distribution of the absolute value of the field in the second component is shown. The solitons displayed in (a) and (b) corresponds to $b_1 = 3$, $b_2 = 5.5$, and $b_1 = 3$, $b_2 = 2.5$, respectively.